\documentclass[useAMS,usenatbib]{mn2e}

\providecommand*{\mnras}{{\it MNRAS}}

\providecommand*{\pre}   {{\it Physical Review E}}
\providecommand*{\apj}   {{\it Astrophys. J.}}
\providecommand*{\apjl}   {{\it Astrophys. J. Lett.}}

\usepackage{natbib}
\title[A Model of Plasma Heating by Large-Scale Flow]{A Model of Plasma Heating by Large-Scale Flow}
\author[P.~Pongkitiwanichakul et al.]{P. Pongkitiwanichakul$^{1}$\thanks{E-mail:peera@oddjob.uchicago.edu}, F.~Cattaneo$^{1}$,
S.~Boldyrev$^{2}$, J.~Mason$^{3}$, and J.C.~Perez$^{4}$ \\
$^{1}$Department of Astronomy \& Astrophysics, University of Chicago, Chicago, IL 60637, USA\\
$^{2}$Department of Physics, University of Wisconsin at Madison,  WI 53706, USA\\
$^{3}$College of Engineering, Mathematics \& Physical Sciences, University of Exeter, Exeter, EX4 4QF, United Kingdom\\
$^{4}$Space Science Center, University of New Hampshire, Durham, NH 03824, USA}

\usepackage{graphicx,color}

\begin{document}

\date{Accepted 1988 December 15. Received 1988 December 14; in original form 1988 October 11}


\maketitle

\label{firstpage}

\begin{abstract}
In this work we study the process of energy dissipation triggered by a slow large scale motion of a magnetized conducting fluid. Our consideration is motivated by the problem of heating the solar corona, which is believed to be governed by fast reconnection events set off by the slow motion of magnetic field lines anchored in the photospheric plasma. To elucidate the physics governing the disruption of the imposed laminar motion and the energy transfer to small scales, we propose a simplified model where the large-scale motion of magnetic field lines is prescribed not at the footpoints but rather imposed volumetrically. As a result, the problem can be treated numerically with an efficient, highly-accurate spectral method, allowing us to use a resolution and statistical ensemble exceeding those of the previous work. We find that, even though the large-scale deformations are slow, they eventually lead to reconnection events that drive a turbulent state at smaller scales. The small-scale turbulence displays many of the universal features of field-guided MHD turbulence like a well developed inertial range spectrum. Based on these observations, we construct a phenomenological model that gives the scalings of the amplitude of the fluctuations and the energy dissipation rate as  functions of the input parameters. We find a good agreement between the numerical results and the predictions of the model.

\end{abstract}


\section{Introduction}
The mechanism by which energy is extracted from large-scale plasma flows and converted into heat is one of the fundamental problems in astrophysical magnetohydrodynamics. The energy may be dissipated in large current and vorticity sheets, transferred to small scales via a turbulent cascade, converted into heat in a very large number of tiny current sheets, or dissipated in a fashion incorporating several of these mechanisms. One of the applications of the theory is the problem of particle heating and acceleration in the solar corona that is hundreds of times hotter than the solar  photosphere~\citep{aschwanden04,klimchuk06}. The most accepted mechanism involves converting coronal magnetic energy into heat. The magnetic field is assumed to be frozen into the plasma, so that the twisting and tangling of magnetic field lines by the photospheric footpoint motions eventually releases magnetic energy in the process of magnetic reconnection~\citep{yamada10}.  

In this study we are interested in the case of a slow large-scale motion of the magnetic field lines, i.e. slow compared to the time scale for the Alfv\'{e}n waves to propagate back and forth along the field lines through the system. This corresponds to the case of slow footpoint motion in coronal heating, where the magnetic loops respond by evolving slowly through a sequence of magnetostatic equilibria~\citep{Gold_Hoyle_1960,Barnes_Sturrock_1972,Ballegooijen_1986}. As not all of the equilibria have the same magnetic topology, the deformation of an equilibrium with one topology into another happens through the process of magnetic reconnection, which converts magnetic energy into kinetic energy of the plasma particles~\citep[e.g.,][]{Parker_1972,Ng_Bha_1998,Priest_Forbes_2000,low2015}.  

This process is often studied in the framework of the Parker  model~\citep{Parker_1972,Parker_1988}, where the coronal loops are treated as straight plasma columns. The initially uniform magnetic field lines in a column are attached at both ends to the perfectly conducting boundaries where the slow plasma motion is prescribed. The problem cannot be solved analytically in its most general form and the available numerical simulations can only address Reynolds numbers that are hopelessly below the astrophysically relevant values.  Consequently, the goal of phenomenological and numerical treatments is to establish the plasma heating rate and, in particular, its scaling with the magnetic Reynolds number $Rm$, in the hope that the results may be extrapolated to the naturally relevant parameter regimes \cite[e.g.,][]{Longcope_Sudan_1994,rappazzo_velli2011,Ng_Lin_Bha_2012,wan2014}. 

In the phenomenological treatment proposed in~\cite{Longcope_Sudan_1994}, it was assumed that in the limit of large $Rm$ the energy dissipation occurs in a finite number of isolated Sweet-Parker current sheets. The heating rate was then predicted to scale as $\epsilon \propto Rm^{1/3}$. In more recent studies by~\cite{Ng_Bha_2008} and \cite{Ng_Lin_Bha_2012} it was proposed that the scaling essentially depends on the rate of magnetic field ``stirring'' by the footpoint motion. If the typical time of the field-line twisting by random footpoint motion is smaller than the typical time of the magnetic energy build-up and release cycle, then the heating rate becomes independent of the Reynolds number.  This conclusion was supported in part by numerical simulations at moderate~$Rm$. 

In this work we propose that  in the limit of large Reynolds numbers, the fast random magnetic line twisting and tangling does not need to be imposed by footpoint motion. Rather, it is naturally provided by the small-scale turbulence that inevitably develops in such a regime.
Once the turbulence is developed, the rate of energy dissipation is dictated by the rate of energy cascade toward small scales, which is independent of the Reynolds number. This picture therefore does not require the random rapid motion of the footpoints. Moreover, it does not require the assumption of a finite number of current sheets in the limit of large~$Rm$. Such an assumption would in fact be incorrect, as the number of current sheets responsible for energy dissipation in MHD turbulence is known to increase as the Reynolds number increases~\cite[e.g.,][]{zhdankin_etal2013}.  

To support our phenomenological theory with numerical simulations, we propose a so-called volumetric Parker model, where instead of prescribing the displacements of the footpoints, we impose a slow large-scale velocity throughout the entire volume of the fluid. The justification for this modification is that for high Alfv\'en velocities a boundary displacement is communicated almost instantaneously along the field lines. A practical advantage of our model is that the volumetric displacement problem can be embedded in a three-dimensional periodic domain and treated numerically with an efficient, highly-accurate spectral method. 
This allows us to study the problem with a resolution and statistical ensemble exceeding those of previous studies, and to derive novel scaling laws for the dependence of the magnetic and kinetic fluctuations on the parameters of the large-scale flow. 

\section{Model}
We solve the full three-dimensional incompressible dissipative MHD equations\footnote{We note that in studies of MHD turbulence with a strong guide field, the reduced MHD approximation is often used. We do not make such an approximation in our analysis.} with an additional volumetric forcing term,
\begin{eqnarray}
  \frac{\partial}{\partial t}{\bf v}+{\bf v}\cdot\nabla{\bf v}-{\bf b}\cdot\nabla{\bf b} -\hat{{\bf v}}_A 
  \cdot\nabla{\bf b} \nonumber\\
  =-\nabla P +S^{-1}\nabla^2 {\bf v}+{\bf F},
  \label{eq:mhd1} \\
   \frac{\partial}{\partial t}{\bf b}+{\bf v}\cdot\nabla{\bf b}-{\bf b}\cdot\nabla{\bf v}-
   \hat{{\bf v}}_A \cdot\nabla{\bf v}  = S^{-1}\nabla^2 {\bf b}, \label{eq:mhd2} \\
 \nabla \cdot {\bf v} =0, \quad  \nabla \cdot {\bf b} =0, 
 \label{eq:mhd3}
  \end{eqnarray}        
where ${\bf
v}_A={\bf B}_0/\sqrt{4\pi \rho_0}$ is the Alfv\'en velocity based upon the uniform
background magnetic field ${\bf B}_0$ which is in the $z$-direction,
${\bf v}$ is the fluctuating plasma velocity normalized by $v_{\rm A}$, ${\bf b}$ is the
fluctuating magnetic field normalized by $v_A$, $P=(p+b^2/2)$, $p$ is the plasma pressure normalized to $\rho_0v_A^2$, $\rho_0$ is the background plasma density, $S=v_AL_\perp/\nu$ is the {(constant)} Lundquist number (the magnetic Prandtl number $Pm$ is set to unity, that is, fluid viscosity is equal to magnetic diffusivity),  where $L_\perp$ is a characteristic scale length transverse to ${\bf B}_0$, and ${\bf F}$ is a forcing term. In these equations time is normalized by the Alfv\'en time $L_\perp/v_A$ and spatial coordinates are normalized by $L_\perp$.

The driving of the system is performed by ensuring that the large-scale velocity field 
${\bf v}_0$, say, that occupies certain wave numbers (see below), is time independent, while the remaining Fourier components of the velocity field are allowed to evolve. 
The prescribed part of the velocity field is defined by 
\begin{eqnarray}
{\bf v}_0 = {\hat {\bf z}}\times{\nabla\psi},
\end{eqnarray}
where
\begin{eqnarray}
\label{eq:stream_f}
& \psi = v_0 L_\perp\sin\left({z}/{L_\parallel}\right)\left[\cos\left({x}/{L_\perp}\right)+\cos\left({y}/{L_\perp}\right)\right]- \nonumber \\ &- \left({v_0 L_\perp}/{2}\right)\cos\left({z}/{L_\parallel}\right)\left[\cos\left({2 x}/{L_\perp}\right)+\cos\left({2 y}/{L_\perp}\right)\right].\quad
\end{eqnarray}
Thus, at each time step, the Fourier components of ${\bf v}_0$ are kept fixed while the remaining components of the velocity are allowed to vary.    

An equivalent way of looking at this driving mechanism is to assume that all of the components of the velocity field are allowed to evolve, but the large-scale force ${\bf F}$ in equation~(\ref{eq:mhd1}) is chosen in such a way that at each time step it brings the large-scale component of the velocity field back to its prescribed value~(\ref{eq:stream_f}). This is the interpretation that we will use in what follows.  

It is important to stress that the imposed weak velocity field~$v_0$ is not strong enough to drive turbulence in the absence of the magnetic field. Its detailed form is not important, however, it should be chosen so that it will engender non-trivial deformations of the magnetic field. Here, we have chosen a flow with a simple, large-scale cellular structure and non-trivial trajectories.
The dynamics of the system thus essentially depends on the build up and release of magnetic energy, which we discuss in detail in the next sections. 

{We also note that we do not aim at providing a detailed explanation of  coronal heating. Rather, we concentrate on a fundamental mechanism of energy extraction from large scale MHD flows. We however believe that in the parameter regime $(v_{\rm A} L_\perp)/(v_0 L_\|) \gg 1$, our approach should be qualitatively similar other approaches based on line-tying.\footnote{We do not make such an approximation here.}} 


The numerical code solves equations (\ref{eq:mhd1}-\ref{eq:mhd3}) in a triply periodic rectangular domain of cross-sectional area $(2\pi L_\perp)^2$ and height $2\pi L_\|$, where the subscripts denote the directions perpendicular and parallel to the background magnetic field. We use a fully dealiased 3D pseudospectral algorithm to perform the spatial discretization on a grid with a resolution of $512^3$ mesh points. A description of the numerical scheme may be found in \cite{cattaneo-ew2003}.

\section{Results}

In the following sections, we present the results of a series of simulations. The parameters of the simulations are summarized in Table~\ref{tab:sim_params}. 
The values of the Lundquist number $S${, which is kept constant in each simulation,} are limited from above by the requirement that the resulting turbulent fluctuations are {well }resolved at the grid scale. {By well resolved we mean that the high wavenumber tail of the spectra has a fast falloff. The values of the Lundquist number $S$} are limited from below by the requirement that the magnetic Reynolds number based on the prescribed  flow $v_0$, $Rm_0=(v_0/v_A)S$ exceeds unity so that the magnetic field lines can be advected by the flow.  
The Reynolds numbers of the fluctuations (which are not known in advance) are calculated after both the velocity and magnetic fields reach statistically steady states.  In some cases, multiple values of $S$ are used to establish the scaling with the Lundquist number. The last column shows the time duration $T$ of each simulation normalized to $L_\perp/v_A$. We also note that larger values of $L_\| v_A/(L_\perp v_0)$ typically mean that a larger computational time is required. {The ratio $v_0/v_{\rm A}$ is quite small in our simulations, which is also the case in the solar corona.}

\begin{table}
\begin{center}
\begin{tabular}{c@{\hspace{0.5cm}}c@{\hspace{0.5cm}}c@{\hspace{0.5cm}}c@{\hspace{0.5cm}}c@{\hspace{0.5cm}}}\hline\hline
Run No. & $L_\parallel/L_\perp$ & $v_0/v_{\rm A}$ & $S$ & $v_{\rm A}T/L_\perp$ \\\hline

A1 & 1 & $1.25\times 10^{-3}$   &  8000 & $1.2\times10^{4}$\\
A2 & 1 & $2.5\times10^{-3}$   &  4000 & $5.8\times10^{3}$\\
A3 & 1 & $5 \times 10^{-3}$   &  4000 & $3.2\times10^{3}$\\
A41 & 1 & $1 \times 10^{-2}$   &  1000 & $5.4\times10^{3}$\\
A42 & 1 & $1 \times 10^{-2}$   &  2000 & $5.4\times10^{3}$\\
A43 & 1 & $1 \times 10^{-2}$   &  4000 & $4.4\times10^{3}$\\
A44 &1 & $1 \times 10^{-2}$   &  8000 & $4.9\times10^{3}$\\
A5 & 1 & $2 \times 10^{-2}$   &  4000 & $1.8\times10^{3}$\\
A6 & 1 & $4 \times10^{-2}$   &  4000 & $1.2\times10^{3}$\\
B1 & 3 & $1.25\times10^{-3}$   &  32000 & $2.0\times10^{4}$\\
B2 & 3 & $2.5 \times10^{-3}$   &  16000 & $1.6\times10^{4}$\\
B3 & 3 & $5 \times10^{-3}$   &  8000 & $2.0\times10^{4}$\\
B41 & 3 & $1 \times10^{-2}$   &  5000 & $6.0\times10^{3}$ \\
B42 & 3 & $1 \times10^{-2}$   &  8000 & $8.4\times10^{3}$ \\
B43 & 3 & $1 \times10^{-2}$   &  12500 & $8.7\times10^{3}$ \\
B5 & 3 & $2 \times10^{-2}$   &  8000 & $5.7\times10^{3}$\\
B6 & 3 & $4 \times10^{-2}$   &  8000 & $1.8\times10^{3}$ \\
C1 & 8 & $1.5 \times10^{-3}$   &  32000 & $9.6\times10^{3}$\\
C2 & 8 & $3.3 \times10^{-3}$   &  32000 & $2.0\times10^{4}$ \\
C3 & 8 & $2 \times10^{-2}$   &  8000 & $9.6\times10^{3}$\\

\hline
\end{tabular}
\caption{\small Summary of the simulation parameters.  Here, $T$ denotes the total physical time for each simulation. 
}\label{tab:sim_params}
\end{center}
\end{table}

\begin{figure}
\includegraphics[width=84mm]{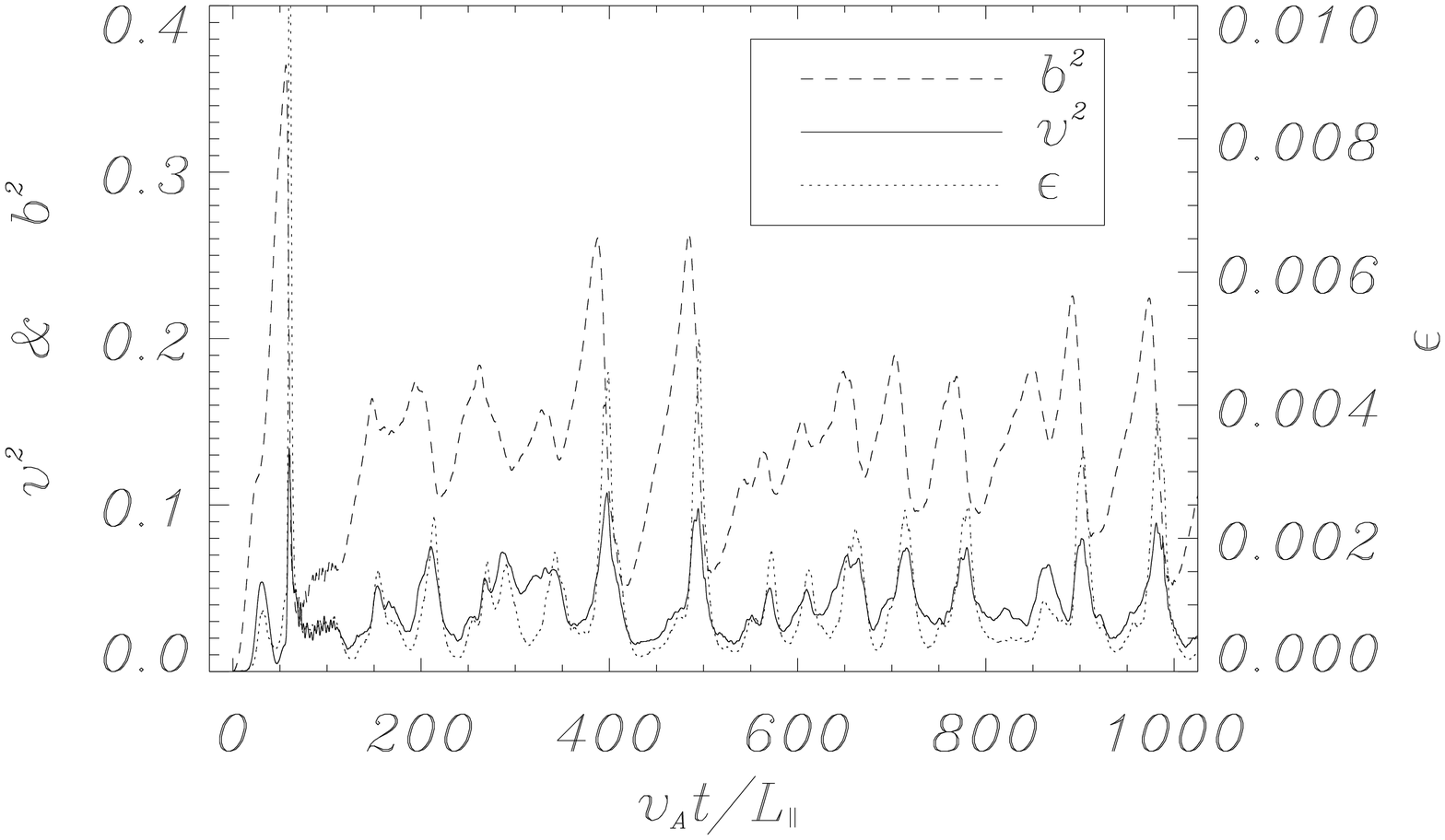}
 \caption{\small The time series of $\epsilon$, $v^2$ and $b^2$ from simulation B42.}
 \label{fig:time_series}
\end{figure}

{\em Cyclic bursts.}---First, we focus on the time evolution of the kinetic and magnetic energies, $v^2$ and $b^2$, and the energy dissipation rate,~$\epsilon$.\footnote{ We identify the dissipation rate with the heating rate~$\epsilon$, which is an appropriate association to make in incompressible MHD as it does not model plasma heating directly.} We do not separate the viscous and ohmic heating as they are similar to each other. As an illustrative example, Figure~\ref{fig:time_series} shows $v^2$, $b^2$ and $\epsilon$ from run~B42. 
The energies and the heating rate are highly intermittent but they are related to each other. At first, the prescribed flow ${\bf v}_0$ disturbs the large scale field ${\bf B}_0$ and the magnetic perturbation ${\bf b}_0$ is generated. The Fourier modes of ${\bf b}_0$ are the same as the modes of ${\bf v}_0$. This process increases the magnetic energy until 
the term  
${\bf b}_0\cdot\nabla {\bf b}_0$ generates the velocity fluctuations~$\delta {\bf v}$ with higher harmonics than those in~${\bf v}_0$. The total velocity field is then ${\bf v}={\bf v}_0+\delta {\bf v}$.

Eventually, the velocity and magnetic perturbations are developed at scales small enough to cause the release of the magnetic energy accumulated in ${\bf b}_0$ by non-ideal processes. This happens in a short time scale in the form of a burst that transfers magnetic energy into plasma flow and heat. This may be  consistent with a fast and intermittent rearrangement of magnetic field lines and energy release due to magnetic reconnection \cite[e.g.,][]{rappazzo2008,fuentes-fernandez2012,osman2012,huang2014,leo13,higash13}.  
After the energy of ${\bf b}_0$ has been released, there is no energy to feed $\delta {\bf v}$. Both ${\bf b}_0$ and $\delta {\bf v}$ then decrease. The energy in ${\bf b}_0$ is then re-accumulated by ${\bf v}_0$ and the process repeats itself. This cyclic process is responsible for the intermittency observed in $v^2$, $b^2$, and $\epsilon$. It is reasonable to assume that the solution is self-correlated during only one cycle, therefore, the cycle period $\tau_c$ also plays the role of the correlation time of the fluctuations in the system. 

\begin{figure}
\includegraphics[width=84mm]{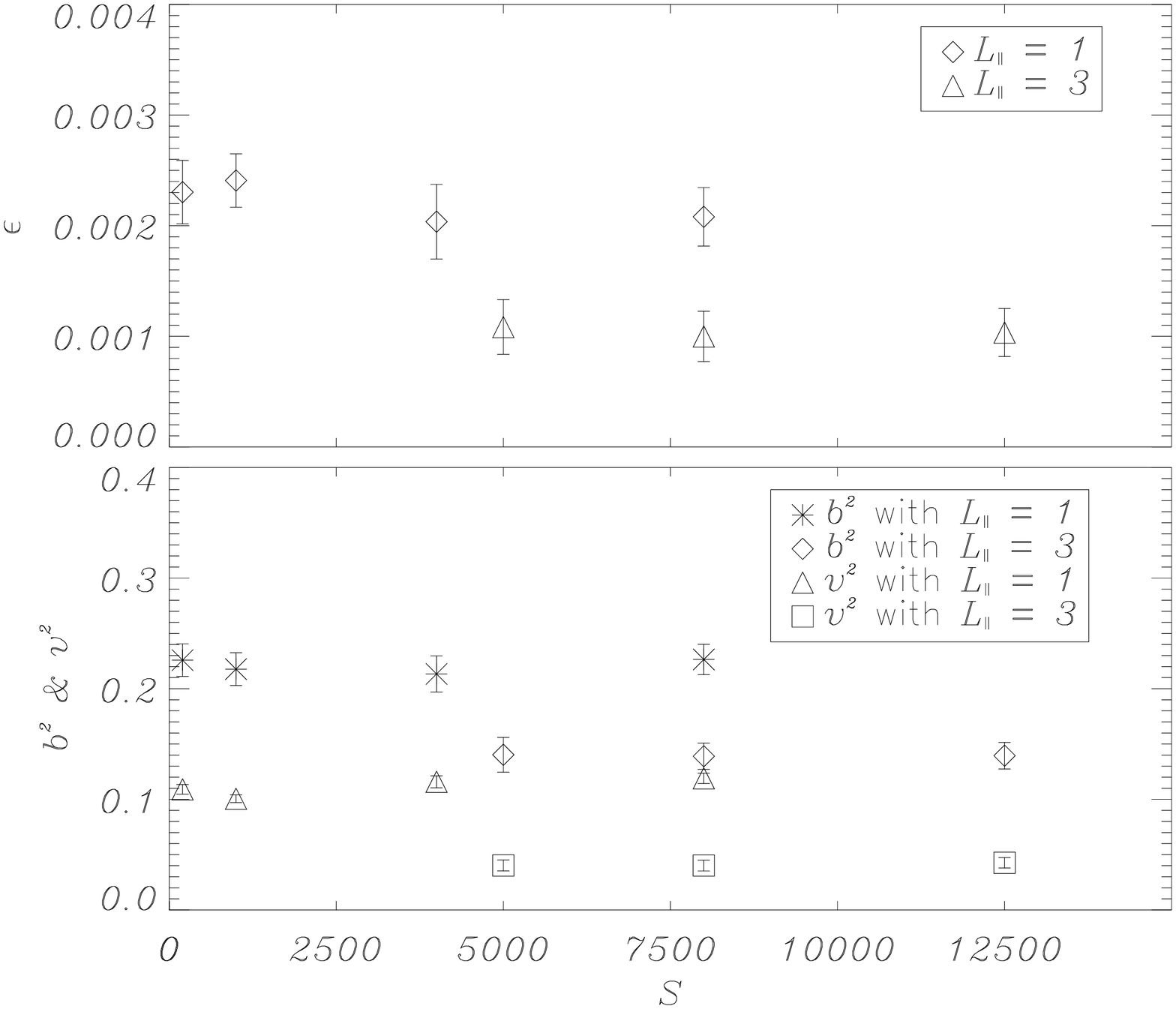}
 \caption{\small The time-averaged values of $\epsilon$ (upper panel) and of $b^2$ and $v^2$ (lower panel) from simulations A41-44 and B41-43. }
\label{fig:heat_s}
\end{figure}

{\em Scaling with Lundquist number.}---To study how $v^2$, $b^2$, and $\epsilon$ depend on the Lundquist number  $S$,
we compare the results from simulations A41-44 and B41-43.  
The time-averaged dissipation rate $\epsilon$, as well as the fluctuation energies, $b^2$ and $v^2$, from these simulations are shown in Figure~\ref{fig:heat_s}. They appear to be independent of~$S$. Since the values of $Re$ are proportional to $S$, we can also claim that $\epsilon$, $b^2$, and $v^2$ are independent of the Reynolds number.

\begin{figure}
\includegraphics[width=84mm]{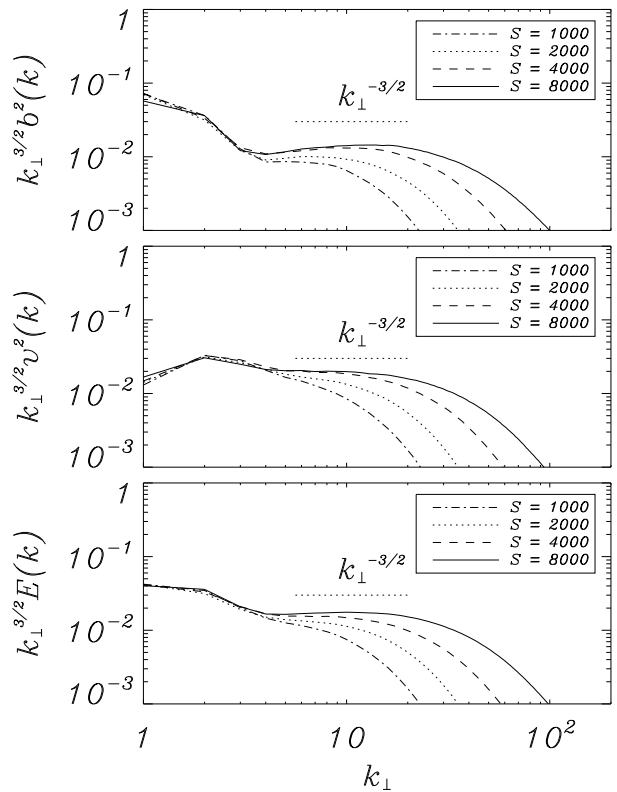}
 \caption{\small The time averaged power spectra of $b^{2}$ (upper panel), $v^2$ (middle panel), and $E=0.5(v^2+b^2)$ (lower panel)  from simulations~A41-44.}
\label{fig:zmp}
\end{figure}

An important feature of our $Pm= 1$ simulations is that both viscous and ohmic heating are very similar.
This is consistent with the presence of small scale turbulence governing the energy cascade and energy dissipation in the system. Indeed, the values of the  time-averaged $Re$ range from  $300$ to $2700$, so turbulence may  develop. To verify this scenario we plot the energy power spectra obtained in simulations A41-44, in Figure~\ref{fig:zmp} . Beyond the transitional range of scales observed at $k_\perp\leq 3$, both magnetic and velocity fluctuations exhibit broad energy spectra with the spectral indices approaching $-3/2$ as $S$ increases. We believe that this is a signature of MHD turbulence developing in the system \cite[e.g.,][]{muller05,tobias_cb11,perez_etal2012}, which is responsible for removing the energy from the large scale and transferring it into heat with the rate independent of the Reynolds and Lundquist numbers.

{\em Scaling with the large-scale parameters.}---In this section, we relate the energies of fluctuations~$v^2$ and $b^2$, and the dissipation rate $\epsilon$ to the large-scale parameters of the system, $B_0$, $v_0$, $L_\parallel$, and $L_\perp$. In steady state turbulence, the rate of energy dissipation at small scales is equal to the rate of energy cascade over scales and to the rate of energy supply at large scales. The driving force in equations (\ref{eq:mhd1}-\ref{eq:mhd3}) is needed to keep the large scale velocity field ${\bf v}_0$ prescribed, therefore, it should balance the large-scale magnetic force, 
estimated as $({{\bf v}}_A\cdot \nabla ) {{\bf b}}_0$.  Recall that the magnetic fluctuations ${\bf b}_0$ result from a shuffling of the uniform magnetic field ${\bf B}_0$ by the large scale flow~${\bf v}_0$, and that  ${\bf b}_0$ has the same set of Fourier modes as~${\bf v}_0$.  The energy input rate per unit volume is therefore estimated as 
\begin{eqnarray}
\epsilon \sim  \langle {\bf v}_0 \cdot {\bf F} \rangle  \sim - \langle {\bf v}_0 \cdot ({{\bf v}}_A\cdot \nabla ) {{\bf b}}_0 \rangle \sim v_A v_0 b_0/L_\|, 
\label{ei}
\end{eqnarray} 
where $L_\|$ is the typical field-parallel scale of~${\bf v}_0$. This rate should coincide with the rate of energy transfer to small scales  in the turbulent cascade,
\begin{eqnarray}
\epsilon 
\sim \langle \delta {\bf v}\cdot (\delta {\bf b}\cdot\nabla_\perp)\delta {\bf b} \rangle \sim \delta v^3/l_\perp,
\label{et} 
\end{eqnarray}  
where $\delta v\sim \delta b$ are the turbulent fluctuations at the outer scale of turbulence, $l_\perp$ (this scale approximately corresponds to $k_\perp = 3$ in Figure~\ref{fig:zmp}). Here the magnetic fluctuations $\delta {\bf b}$ have Fourier harmonics with $k\geq 3$. 

In a steady state, there is a balance between generation of the large-scale magnetic fluctuations ${\bf b}_0$, and their diffusion due to small-scale turbulence. The generation of magnetic fluctuations by the large-scale flow ${\bf v}_0$ is given by the term $({\bf v}_A\cdot \nabla){\bf v}_0$ with the magnitude $\sim v_A v_0/L_\|$ in the induction equation, while their diffusion due to turbulence is described by $\eta_T b_0/L_\perp^2\sim l_\perp \delta v b_0/L_\perp^2$, where we have substituted $\eta_T \sim l_\perp \delta v$ for the turbulent diffusivity and $L_\perp$ is the typical field-perpendicular scale of~${\bf v}_0$. We thus arrive at the balance condition
\begin{eqnarray}
v_Av_0/L_\| \sim l_\perp \delta v b_0/L_\perp^2.
\label{diff}
\end{eqnarray}

\begin{figure}
\includegraphics[width=84mm]{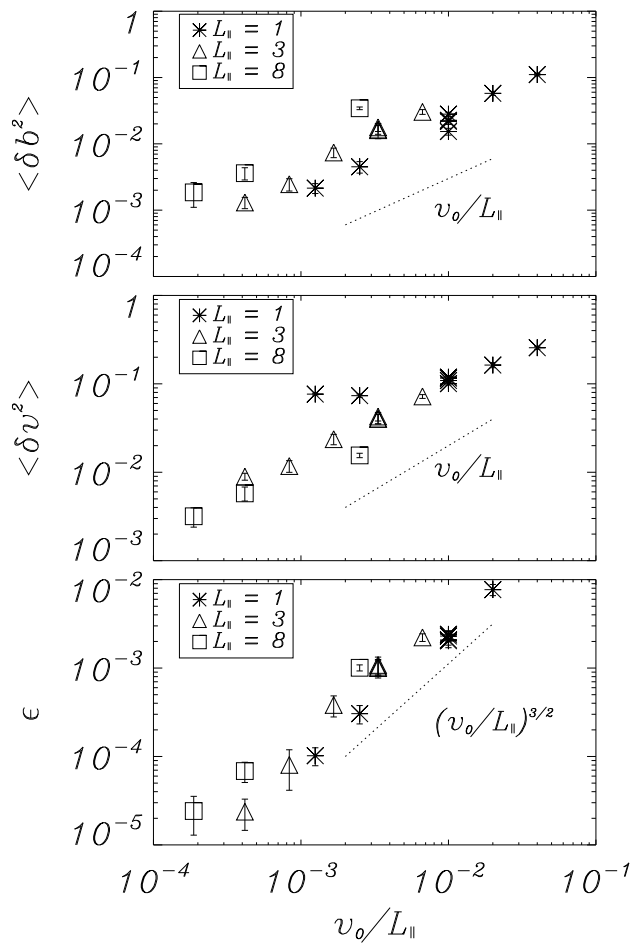}
 \caption{\small The time averaged $\delta b^2$ at $k\geq 3$ , $\delta v^2$, and $\epsilon$ from all simulations versus $v_0/L_\parallel$. (The slight upshift of the magnetic fluctuations, denoted by squares in the upper panel, may be related to our somewhat subjective separation of the stronger large-scale component~$b_0$ from the weaker turbulent component~$\delta b$ at~$k=3$. No upshift is observed in the corresponding velocity fluctuations that do not develop a strong large-scale component.)}
\label{fig:dx3}
\end{figure}

From (\ref{ei}), (\ref{et}), and (\ref{diff}) we obtain the amplitudes of turbulent fluctuations 
\begin{eqnarray}
\delta v^2\sim \delta b^2 \sim  v_0 v_AL_\perp/{L_\|},
\label{turb_fields_scaling}
\end{eqnarray}
and the energy dissipation rate per unit mass 
\begin{eqnarray}
\epsilon \sim \left({v_0 v_{\rm A}}L_\perp/{L_\|}\right)^{3/2}l_\perp^{-1}.
\label{diss_scaling}
\end{eqnarray}
Figure~\ref{fig:dx3} shows how $\delta b^2$, $\delta v^2$ and $\epsilon$ depend on $v_0/L_\parallel$ in all of the simulations. They are in good agreement with our phenomenological estimates. We also note that from equations (\ref{ei}), (\ref{et}), and (\ref{diff}) one derives $(\delta b/b_0)^2\sim (l_\perp/L_\perp)^2$, which is  consistent with Figure~\ref{fig:zmp}. 

Finally, we are in a position to estimate the large-scale correlation time~$\tau_c$. Since large-scale magnetic fluctuations grow up to $b_0$ on the correlation time scale, from the induction equation we obtain ${b_0}/\tau_c \sim {v_A}{v_0}/L_\|$. Substituting for $b_0\sim (L_\perp/l_\perp)\delta b$, and using result~(\ref{turb_fields_scaling}), we estimate the correlation time 
\begin{eqnarray}    
\tau_c\sim (L_\|/v_A)^{1/2}(L_\perp/v_0)^{1/2}(L_\perp/l_\perp).
\end{eqnarray}
According to our previous discussion, this time also characterizes the burst cycles of large-scale fluctuations.  

\section{Discussion and Conclusion}
We have presented a numerical and phenomenological study of the coronal heating problem based on the so-called volumetric Parker model. In this model the slow large-scale motion of the magnetic field lines is prescribed not at the boundaries of the domain, but throughout the volume of the fluid. As a result, the model allows for an effective numerical study, with the resolution, statistical ensemble, and accuracy exceeding those of previous treatments. 

We have established that the energy is extracted from the prescribed slow flow in a bursty, intermittent fashion. Energy is quickly redistributed over large scales and transferred to small dissipative scales (large $k_\perp$ modes) through a universal turbulent cascade. The energy dissipation rate and the levels of the magnetic and velocity fluctuations are independent of the magnetic Reynolds and Lundquist numbers. However, there is a dependence on the large-scale parameters of the prescribed flow, with the scaling given by equations~(\ref{turb_fields_scaling}) and (\ref{diss_scaling}).

To give a sense of magnitudes and scales, we make order of magnitude estimates based on equation~(\ref{diss_scaling}).  If a proton-electron plasma is assumed, then equation~(\ref{diss_scaling}) can be used to estimate the energy dissipation rate per unit volume
\begin{eqnarray}
& E_{\rm diss} = \epsilon \rho = 5.3\times 10^{-3}\left(\frac{v_0}{10^{5} {\rm cm/s}}\right)^{3/2} 
\left(\frac{v_{\rm A}}{10^{8} {\rm cm/s}}\right)^{3/2}\nonumber \\
&\times \left(\frac{n_0}{10^{9} {\rm cm^{-3}}}\right)\left(\frac{L_\perp}{L_\parallel}\right)^{3/2}\left(\frac{l_\perp}{10^7 cm}\right)^{-1} {\rm erg\, s^{-1} cm^{-3}}, 
\label{diss_withunit}
\end{eqnarray}
where $n_0$ is the number of protons or electrons, and $\rho$ is the mass density. If we assume {$v_0 = 5\times10^5$ cm/s}, $v_{\rm A} = 2\times 10^8$ cm/s, {$n_0 = 2.5\times10^9$ cm$^{-3}$,} $L_\perp = 10^8$ cm, $L_\parallel = 4\times 10^9$ cm and $l_\perp = L_\perp/2$, the dissipation rate becomes {$E_{\rm diss}=3.33\times10^{-4}$ erg s$^{-1}$ cm$^{-3}$.} This rate is equivalent to having the energy flux {$E_{\rm diss} L_\parallel= 1.32\times 10^6$ erg s$^{-1}$ cm$^{-2}$} through the boundaries with an area of $L^2_\perp$. {This energy flux is within order of magnitude agreement with the required flux to heat the corona \citep[cf.][]{aschwanden04,depontieu07,depontieu11,mcintosh11}. This estimate for the energy flux is not unreasonble given the idealized nature of our model.} For an extensive review of the most recent advances in understanding the coronal heating problem and the acceleration of the solar wind, the reader is referred to~\citet{demoortel15}.

Our treatment is different and complementary to the previous studies where the role of imposed random tangling of magnetic field lines and magnetic turbulence in the energy dissipation was also discussed  \cite[e.g.,][]{Dmitruk98,Dmitruk99,rappazzo2008,Ng_Lin_Bha_2012}. In those studies the magnetic field columns were distorted at the boundaries rather than volumetrically, which mathematically imposed perturbations with a broad spectrum of field parallel wave numbers~$k_\|$. In our case, the driving is strictly confined to the large-scale modes so that the small-scale dissipation cannot be affected by the driving directly. The observed spectrum and scaling of the fluctuations is solely a consequence of the nonlinear dynamics.   In addition, in contrast with \cite{Ng_Lin_Bha_2012}, we observed the $S$-independent dissipation rate without imposing an external fast field-line shuffling. In contrast with \cite{rappazzo2008}, we use the full MHD treatment (rather than RMHD) as we do not make a priori assumptions about the nature of turbulence that may develop in the system (e.g., the relevance of field-parallel fluctuations, transport of field-parallel momentum, etc., which are all neglected in RMHD). Finally, in contrast with \cite{Dmitruk98} and \cite{Dmitruk99} where the 2D MHD equations were used, we use a full 3D treatment and our analytically and numerically derived scaling of the fluctuating fields (\ref{turb_fields_scaling}) is different from their predictions. 

\section*{Acknowledgments}

This research was supported by the NSF Center for Magnetic
Self-Organization in Laboratory and Astrophysical Plasmas at
the University of Chicago, by the US DOE award
no. DE-SC0003888, by the NASA grant no. NNX11AE12G, and by the National Science Foundation under
grant no. NSF PHY11-25915 \& no. AGS-1261659. SB and JM appreciate
the hospitality and support of the Kavli Institute for Theoretical
Physics, University of California, Santa Barbara,
where part of this work was conducted. Simulations were performed at the Texas Advanced Computing
Center (TACC) at the University of Texas at Austin under
the NSF-Teragrid Projects TG-AST140015 \& TG-PHY120042 and by the
National Institute for Computational Sciences.

\providecommand{\noopsort}[1]{}\providecommand{\singleletter}[1]{#1}%

\end{document}